\documentclass[12pt]{iopart}
\include{graphicx}

\newcommand{\rsun}{\ensuremath{{\rm R}_{\odot}}}

\newcommand{\rhov}{\ensuremath{\rho \left| {\mathbf{v}_p} \right|}}
\newcommand{\bvec}{\ensuremath{\mathbf{B}}} 

\def\lesssim{\mathrel{\hbox{\rlap{\hbox{\lower4pt\hbox{$\sim$}}}\hbox{$<$}}}}
\def\gtrsim{\mathrel{\hbox{\rlap{\hbox{\lower4pt\hbox{$\sim$}}}\hbox{$>$}}}} 

\begin{document}

\title[]{Magnetically Driven and Collimated Jets  from the Disc-Magnetosphere Boundary of Rotating Stars}

\author{R.V.E. Lovelace, M.M. Romanova, \& P. Lii}
  \address{Department of Astronomy, Cornell University, Ithaca, N.Y. 14853, 
  USA; RVL1@cornell.edu}

\ead{lovelace@astro.cornell.edu}

\begin{abstract}
   We discuss recent progress in understanding  the launching of outflows/jets from the disc-magnetosphere boundary of slowly and 
rapidly rotating magnetized stars.   
    In most of the discussed models the  interior of the
disc  is assumed to have a turbulent viscosity and
magnetic diffusivity (as described
by two ``alpha'' parameters), whereas the coronal region outside
of the disc is treated using ideal magnetohydrodynamics (MHD).
     Extensive MHD simulations have established the occurrence of
 long-lasting outflows  in both the cases of slowly and
 rapidly rotating stars.
({\bf 1})   In the case of {\it slowly rotating stars}, 
a new type of outflow, {\it a conical wind},  is found and studied in simulations.
  The conical winds appear in cases where 
the magnetic flux of the star is bunched up by the inward motion  of the
accretion disc.
      Near their region of origin, the winds have
the shape of a thin conical shell with a half-opening angle of $ \sim 30^\circ$.
   At large distances these outflows become magnetically collimated by
their toroidal magnetic field and form  matter dominated jets.  That is, the
jets are current carrying.
      About $10-30\%$ of the disc matter  flows from the inner disc into the conical winds.
The conical winds may be responsible for episodic  as well as long-lasting
outflows in different types of stars.    The predominant driving force for the 
conical winds is the magnetic force proportional to the negative gradient
of the square of the toroidal magnetic field and not the centrifugal force.
        ({\bf 2}) In the case of {\it rapidly rotating stars} in the ``propeller regime,'' 
 a two-component outflow is observed.
   The first component is similar to the matter dominated conical winds. A large fraction of 
the disc matter may be ejected into the winds in this regime.
     The second component is a high-velocity, low-density
magnetically dominated {\it axial jet} where matter flows along the opened polar 
field lines of the star.
  The axial jet has a mass flux about $10\%$ that of
the conical wind, but its energy flux due to the Poynting flux
 can be larger than the energy flux of the conical wind.
     The jet's angular momentum flux is carried by the magnetic field and
 causes the star to spin-down rapidly.
Such propeller-driven outflows may be responsible for the jets 
in protostars and for their rapid spin-down. 

      When the artificial requirement of symmetry about the 
equatorial plane is dropped, we show that  the conical winds may 
alternately come from one side of the disc and then the other
even for the case where the stellar magnetic field is a
centered axisymmetric dipole.

    Recent MHD simulations of disc accretion
to rotating stars in the propeller regime have been done with
{\it no} turbulent viscosity and {\it no} diffusivity.
    The strong turbulence we observe is due to the magneto-rotational
instability.  This turbulence drives  accretion in the disc and
leads to episodic outflows.

\end{abstract}

\maketitle

\section{Introduction}

Outflows in the form of jets and winds are observed from many disc accreting 
objects ranging from young stars to systems with white 
dwarfs, neutron stars, and black holes.
   A large body of observations exists for outflows
from young stars at different stages of their 
evolution, ranging from protostars, where powerful
collimated outflows -  jets -  are observed, to classical T Tauri stars (CTTSs) where the outflows are weaker and often  less collimated (see review by Ray et al. 2007).
Correlation between the disc's radiated power and the jet power has been found in many CTTSs (Cabrit et al. 1990;
 Hartigan, Edwards \& Gandhour 1995).
A significant number of CTTSs show signs 
of outflows in spectral lines, in particular in
He I where two distinct components of outflows 
had been found (Edwards et al. 2003, 2006; Kwan, Edwards, \& Fischer 2007).
 Outflows are also observed from accreting compact stars such as accreting white dwarfs in symbiotic binaries (Sokoloski \& Kenyon 2003),
or from the vicinity of neutron stars, such as from  Circinus X-1 (Heinz et al. 2007).

Different theoretical models have been proposed 
to explain the outflows from protostars and CTTSs
(see review by Ferreira, Dougados, \& Cabrit 2006).
The models include those where the outflow 
originates from a radially distributed disc
wind (Blandford \& Payne 1982; K\"onigl \& Pudritz 2000;
Casse \& Keppens 2004; Ferreira et al. 2006) or  
from the innermost region of the accretion disc (Lovelace, Berk \& Contopoulos 1991).    The latter model is related to the
 X-wind  model (Shu et al. 1994; 2007; Najita \& Shu 1994; Cai et al. 2008) where 
the outflow originates from the vicinity of the disc-magnetosphere boundary.   
Progress in understanding the theoretical models
has come from MHD simulations of accretion discs around rotating
magnetized stars as discussed below.    Laboratory experiments are
also providing insights into jet formation processes (Hsu \& Bellan 2002;
Lebedev et al. 2005) but these are not discussed here.

Outflows or jets from the disc-magnetosphere boundary were found in early axisymmetric MHD simulations by
Hayashi, Shibata \& Matsumoto (1996) and  Miller \& Stone (1997). 
A one-time episode
of outflows from the inner disc and inflation of the 
innermost field lines connecting the star and the disc were observed
for a few dynamical time-scales.
        Somewhat longer simulation runs were performed 
 by Goodson et al. (1997, 1999), Hirose et al. (1997),
Matt et al. (2002) and K\"uker, Henning \& R\"udiger (2003) 
where several episodes of field inflation and outflows were observed.
These simulations hinted at a possible long-term nature for the outflows.
 However,  the simulations were not sufficiently long to establish
the behavior of the outflows.
     MHD simulations showing long-lasting (thousand of
orbits of the inner disc)  outflows from the disc-magnetosphere have been obtained by our group
 (Romanova et al. 2009;  Lii et al. 2012).
   We obtained these outflows/jets
in two main cases: (1) where the star rotates slowly but the field lines are bunched up into an X-type configuration,  and (2) where
the star rotates rapidly, in the ``propeller regime''
(Illarionov \& Sunyaev 1975;
Alpar \& Shaham 1985; Lovelace, Romanova \& Bisnovatyi-Kogan 1999) .
    Figure 1 shows the equatorial angular rotation rate $\Omega(r,z=0)$
of the plasma in the two cases.   Here,  $r_*$ is the radius
of the star;  $r_{\rm m}$ is the magnetospheric radius where the
kinetic energy density of the disc matter is about equal to the   energy
density of the magnetic field; and 
$r_{\rm cr}= (GM/\Omega_*^2)^{1/3}$
is the co-rotation radius where the angular rotation rate of the
star $\Omega_*$ equals that of the Keplerian disc $\Omega_K=
(GM/r^3)^{1/2}$.    For a slowly rotating star $r_{\rm m} < r_{\rm cr}$
whereas for a rapidly rotating star in the propeller regime
$r_{\rm m} >r_{\rm cr}$.

\begin{figure}
\begin{center}
  \includegraphics[width=5in]{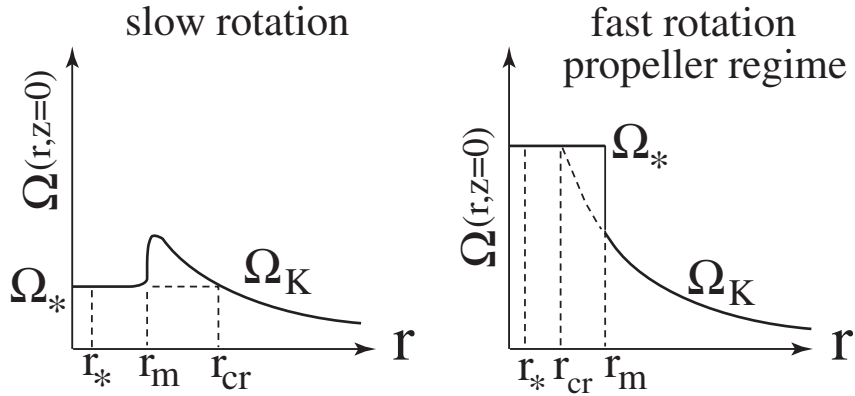}
  \caption{Schematic profiles of the midplane angular velocity
  of the plasma for the case of a slowly rotating star (left-hand panel)
  and a rapidly rotating star (right-hand panel) which is in the
  propeller regime.  Here, $\Omega_*$ is the angular
  rotation rate of the star, $\Omega_K$ is the Keplerian rotation
  rate of the disc,  $r_*$ is the star's radius, $r_{\rm m}$ is the
  radius of the magnetosphere, and $r_{\rm cr}$ is the co-rotation
  radius.}
  \end{center}
\end{figure}
       
       Figure 2 shows examples of the outflows in the
two cases.
In both cases, two-component outflows are observed:
One component originates at the inner edge 
of the disc  near $r_{\rm m}$ and has a narrow-shell conical shape close to the disc and therefore is termed a ``conical wind".   It is matter dominated
but can become collimated at large distances due to its toroidal magnetic
field.
The other component is a magnetically dominated high-velocity
``axial jet'' which flows along the open stellar magnetic field lines. 
             The axial jet may be very strong in the propeller regime.
  A detailed discussion of the simulations and analysis can
be found in Romanova et al. (2009) and Lii et al. (2012).

     Sec. 2 describes the simulations.   Sec. 3 discusses the conical winds and axial jets, the driving and collimation forces, and the variability of
the winds and jets.  Sec. 4 discusses simulation 
results on one-sided and lop-sided jets.  Sec. 5 gives the conclusions.

\begin{figure}
\begin{center}
  \includegraphics[width=5.in]{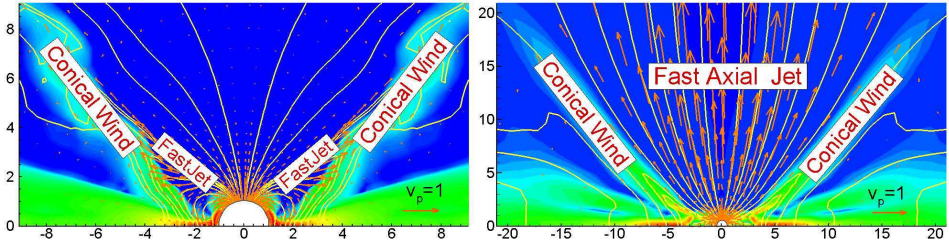}
  \caption{Two-component outflows observed in slowly (left) and rapidly (right) rotating magnetized stars adapted from Romanova et al. (2009).
The background shows
the poloidal matter flux $F_m=\rho v_p$, the arrows are the  poloidal 
velocity vectors, and the lines are sample magnetic field lines. 
The labels point to the main outflow components.}
\end{center}
\end{figure}

\section{MHD Simulations}

We simulate the outflows resulting from disc-magnetosphere interaction
using the equations of axisymmetric MHD.

Outside of the disc the flow is described by the
equations of ideal MHD.
     Inside the disc the flow is described by
the equations of viscous, resistive MHD.
In an inertial reference frame the equations are:
\begin{equation}\label{eq1}
\displaystyle{ \frac{\partial \rho}{\partial t} + {\bf \nabla}\cdot
\left( \rho
{\bf v} \right)} = 0~,
\end{equation}
\begin{equation}\label{eq2}
{\frac{\partial (\rho {\bf v})}{\partial t} + {\bf
\nabla}\cdot
{\cal T} } = \rho ~{\bf g}~,
\end{equation}
\begin{equation}\label{eq3}
{\frac{\partial {\bf B}}{\partial t} - {\bf
\nabla}\times ({\bf v} \times {\bf B}) + {\bf \nabla} \times\left(
\eta_t {\bf \nabla}\times {\bf B} \right)} = 0~,
\end{equation}
\begin{equation}\label{eq4}
{\frac{\partial (\rho S)}{\partial t} + {\bf
\nabla}\cdot ( \rho S {\bf v} )} =  Q~.\
\end{equation}
 Here, $\rho$ is the density,  $S$ is the specific entropy, $\bf
v$ is the flow velocity, $\bf B$ is the magnetic field, $\eta_t$
is the  magnetic diffusivity,  $\cal{T}$ is the
momentum flux-density tensor, $Q$ is
the rate of change of entropy per unit volume,  and ${\bf g} = - (GM /r^{2})\hat{{\bf r}}$
is the gravitational acceleration due to the star, which has mass $M$. 
The total mass of the disc is assumed to
be negligible compared to $M$.  
    Here, ${\cal T}$ is the sum of the ideal plasma terms {\it and}
the $\alpha-$viscosity terms discussed in the next paragraph.
     The  plasma is considered to be an
ideal gas with adiabatic index $\gamma =5/3$, and $S=\ln(p/
\rho^{\gamma})$. We use spherical coordinates $(r, \theta, \phi)$ with  $\theta$ measured from the symmetry axis.  
 The  equations in spherical coordinates are given in
Ustyugova et al. (2006).

     Both the viscosity
and the magnetic diffusivity of the disc plasma are considered
to be due to turbulent fluctuations of the velocity and the magnetic field.
Both effects are non-zero only inside the disc as determined by
a density threshold.
       The  microscopic transport coefficients are  replaced by
 turbulent coefficients.
       The values of these coefficients are assumed to be given by
the  $\alpha$-model of Shakura and Sunyaev (1973), where the
coefficient of the turbulent kinematic viscosity is $\nu_t =
\alpha_\nu c_s^{2}/\Omega_K$, where $c_s$ is the isothermal sound
speed and $\Omega_K(r)$ is the Keplerian angular velocity.  
We take into account the viscous stress terms ${\cal T}_{r\phi}^{\rm vis}$
and ${\cal T}_{\theta \phi}^{\rm vis}$ (Lii et al. 2012).
        Similarly, the coefficient of the turbulent magnetic
diffusivity $\eta_t=\alpha_\eta c_s^{2}/\Omega_K$. Here,
      $\alpha_\nu$ and $\alpha_\eta$ are dimensionless coefficients which
are treated as parameters of the model.

The MHD equations are solved in
dimensionless form so that the results can be readily applied
to different accreting stars.

\begin{table*}
\centering
\caption{Reference values for different types of stars. We choose the mass $M$, radius $R_*$,
equatorial magnetic field $B_*$ and  the period $P_*$ of the star and derive the other reference values.
   The reference
mass $M_0$ is taken to be the mass $M$ of the star.
  The reference radius is taken to be twice the radius
of the star, $R_0=2 R_*$.
    The surface
magnetic field $B_*$ is
different for different types of stars.
  The reference velocity is $v_0=(GM/R_0)^{1/2}$.
The reference time-scale $t_0=R_0/v_0$, and the reference angular velocity $\Omega_0=1/t_0$.
    We measure time
in units of $P_0=2\pi t_0$ (which is the Keplerian rotation period at $r=R_0$).
In the plots we  use the dimensionless time $T=t/P_0$.
    The reference magnetic field is $B_0=B_*(R_*/R_0)^{3}/{\tilde\mu}$,
where $\tilde\mu$ is the dimensionless magnetic moment which has
a numerical value of $10$ in the simulations discussed here.
    The reference density is taken to
be  $\rho_0 = B_0^{2}/v_0^{2}$.
The reference pressure is $p_0=B_0^{2}$.
The reference temperature is
$T_0=p_0/{\cal R} \rho_0 = v_0^{2}/{\cal R}$, where ${\cal R}$ is the gas constant.
   The reference accretion rate is $\dot M_0
= \rho_0 v_0 R_0^{2}$.
   The reference energy flux is $\dot
E_0=\dot M_0 v_0^{2}$.
The reference angular momentum flux is $\dot{L}_0=\dot M_0 v_0
R_0$.   The poloidal  magnetic field
of the star (in the absence of external plasma) is an aligned dipole field.
} 
\begin{tabular}{l@{\extracolsep{0.2em}}l@{}lllll}

\hline
&                                              & Protostars      & CTTSs        & Brown dwarfs   & White dwarfs        & Neutron stars    \\
\hline

\multicolumn{2}{l}{$M(M_\odot)$}                  & 0.8            & 0.8            & 0.056             & 1                   & 1.4       \\
\multicolumn{2}{l}{$R_*$}                         & $2R_\odot$     & $2R_\odot$     & $0.1R_\odot$       & 5000 km             & 10 km     \\
\multicolumn{2}{l}{$R_0$ (cm)}                    & $2.8\cdot10^{11}$    & $2.8\cdot10^{11}$    & $1.4\cdot10^{10}$      & $10^9$            & $2\cdot10^6$    \\
\multicolumn{2}{l}{$v_0$ (cm s$^{-1}$)}           & $1.95\cdot10^7$      & $1.95\cdot10^7$      & $1.6\cdot10^7$          & $3.6\cdot10^8$   & $9.7\cdot10^{9}$ \\
\multicolumn{2}{l}{$P_*$}                         & $1.04$ days    & $5.6$ days     & $0.13$ days        & $89$ s              & $6.7$ ms     \\
\multicolumn{2}{l}{$P_0$}                         & $1.04$ days    & $1.04$ days    & $0.05$ days       & $17.2$ s            & $1.3$ ms   \\
\multicolumn{2}{l}{$B_*$ (G)}                      & $3.0\cdot10^3$      & $10^{3}$      & $2\cdot10^{3}$            & $10^{6}$            & $10^{9}$    \\
\multicolumn{2}{l}{$B_0$ (G)}                     & 37.5           & 12.5          & 25.0                & $1.2\cdot10^4$            & $1.2\cdot10^{7}$  \\
\multicolumn{2}{l}{$\rho_0$ (g cm$^{-3}$)}         & $3.7\cdot10^{-12}$  & $4.1\cdot10^{-13}$  & $1.4\cdot10^{-12}$        & $1.2\cdot10^{-9}$         & $1.7\cdot10^{-6}$  \\
\multicolumn{2}{l}{$n_0$ (cm$^{-3}$)}            & $2.2\cdot10^{12}$   & $2.4\cdot10^{11}$   & $8.5\cdot10^{11}$         & $7\cdot10^{14}$         & $10^{18}$          \\
\multicolumn{2}{l}{$\dot M_0$($M_\odot$yr$^{-1}$)}  & $1.8\cdot10^{-7}$  & $2\cdot10^{-8}$   & $1.8\cdot10^{-10}$      & $1.3\cdot10^{-8}$         & $2\cdot10^{-9}$  \\
\multicolumn{2}{l}{$\dot E_0$ (erg s$^{-1}$)}       & $2.1\cdot10^{33}$  & $2.4\cdot10^{32}$   & $2.5\cdot10^{30}$      & $5.7\cdot10^{34}$         & $6\cdot10^{36}$  \\
\multicolumn{2}{l}{$\dot L_0$ (erg s$^{-1}$)}       & $3.1\cdot10^{37}$  & $3.4\cdot10^{36}$   & $1.7\cdot10^{33}$      & $1.6\cdot10^{35}$         & $1.2\cdot10^{33}$  \\
\multicolumn{2}{l}{$T_d$ (K)}                       & $2290$       & $4590$        & $5270$            & $1.6\cdot10^{6}$          & $1.1\cdot10^{9}$  \\
\multicolumn{2}{l}{$T_c$ (K)}                       & $2.3\cdot10^{6}$   & $4.6\cdot 10^{6}$    & $5.3\cdot10^{6}$        & $8\cdot10^{8}$          & $5.6\cdot10^{11}$  \\
\hline
\end{tabular}
\end{table*}

The system of MHD equations (1-4) have been 
 integrated numerically in spherical $(r,\theta,\phi)$ coordinates
  using a Godunov-type numerical scheme.
The calculations were done
in the region $R_{\rm in} \leq r \leq R_{\rm out}$, $0\leq \theta
\leq \pi/2$. The grid is uniform in the $\theta$-direction with $N_\theta$ cells.
    The $N_r$ cells in the radial direction  have $ dr_{j + 1} = (1 + 0.0523)dr_j$ ($j=1..N_r$) so
 that the poloidal-plane cells are curvilinear rectangles with approximately
equal sides.    This choice results in high spatial resolution near the star 
where the disc-magnetosphere interaction takes place while 
also permitting a large simulation region. 
    We have used a range of resolutions going from
 $N_r\times N_\theta = 51\times 31$  to $121\times 51$.
 
\begin{figure*}
\centering
\includegraphics[width=5in]{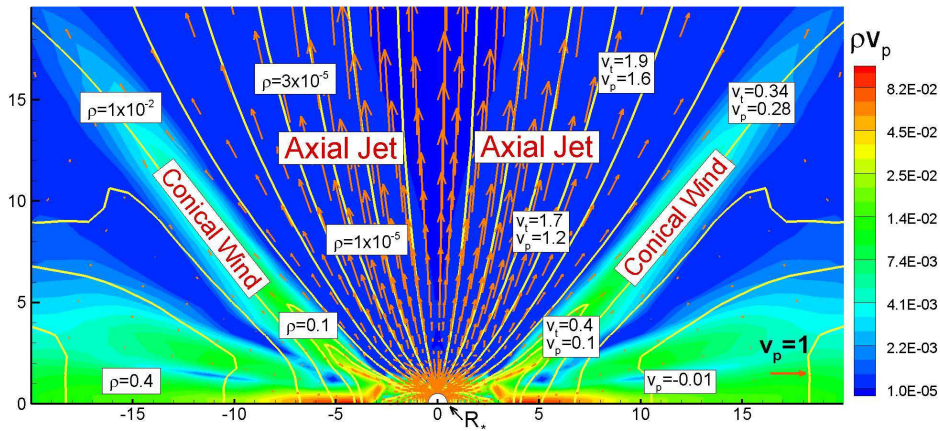}
\caption{Matter flux $\rho v_p$ (background), sample field lines, and poloidal velocity vectors in a propeller-driven outflow at time $t=1400$ (rotations at $r=1$) adapted from Romanova et al. (2009).    
Sample numerical values are given for the poloidal $v_p$ and  total $v_t$ velocity, and for the density $\rho$ for different parts of the simulation region.
One can see from the Table 1 that for CTTSs $v_p=1$ corresponds to $v_p=195$ km/s in dimensional units.
   Unit density corresponds to $\rho_0=4.1\times 10^{-13}$ g cm$^{-3}$. Dimensionless data shown on the plot can be converted to dimensional units for other types of stars  using the reference values from the Table 1.}
\end{figure*}

\section{Conical Winds and  Axial Jets}

      A large number of simulations were done in order
to understand  the origin and nature of conical winds.
  All of the key parameters were varied in order   to ensure that
 there is no special  dependence on any parameter.
  We observed that the formation of conical winds
is a common phenomenon for a wide range of parameters.
They are most persistent and strong in cases where 
the  viscosity and diffusivity coefficients are not very 
small, $\alpha_\nu\gtrsim 0.03$, $\alpha_\eta \gtrsim 0.03$.
Another important condition is  that $\alpha_\nu \gtrsim\alpha_\eta$;
that is, the magnetic Prandtl number
of the turbulence, ${\cal P}_m=\alpha_\nu/\alpha_\eta \gtrsim 1$.
This condition favors the bunching of the stellar magnetic
field by the accretion flow.

       Figure 3 shows a  snapshot from our simulations at time $t=1400$ for  the propeller regime.
The figure shows the dimensionless density and velocity at sample points.
One can see that the velocities in the conical wind component are similar to those in conical winds around slowly rotating stars.
  Matter launched from the disc initially has
an  approximately Keplerian  azimuthal velocity, $v_K=\sqrt{GM_*/r}$.
   It is gradually accelerated to poloidal velocities $v_p\sim (0.3-0.5) v_K$
and the azimuthal velocity decreases.
The flow has a high density and carries most of the disc mass into the outflows. 
The situation is the opposite in the axial jet component where the density is $10^2-10^3$ times lower, while the poloidal and total velocities are
significantly  higher.
   Thus we find a  {\it two-component outflow}:  a matter
 dominated conical wind and a magnetically dominated axial jet.

\begin{figure*}
\centering
\includegraphics[width=5in]{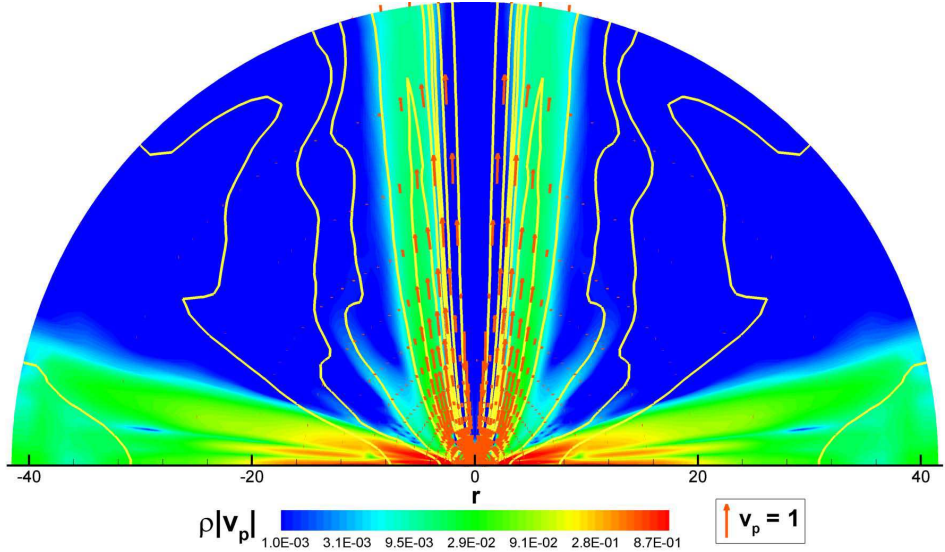}
\caption{The conical wind/jet from a slowly rotating star
at time $t= 860$ adapted from Lii et al. (2012).
The background shows the poloidal matter flux density $\rho{\bf v}_p$ and the lines show 
the poloidal projections of the magnetic magnetic field.
The red vectors show the poloidal matter velocity ${\bf v}_p$.
Dimensional values can be obtained from Table 1. 
For example for a CTTS, $t_0 = 0.366$ days, $R_0= 2\rsun$,  $ t=860$ corresponds
to $315$ days,  and the simulation region is $0.39$ AU in radius. 
The horizontal axis shows the distance  from the star in units of the reference radii $R_0$.  For this case $\alpha_\nu=0.3$ and $\alpha_\eta=0.1$. }
\end{figure*}

   We observe conical winds in both slowly and rapidly rotating stars.
In both cases, matter in the conical winds passes through the Alfv\'en surface
(and shortly thereafter through the fast magnetosonic point),
beyond which the flow is matter-dominated in the sense that the energy flow is carried mainly by the matter. 
    The situation is different for the axial jet component where the flow
is sub-Alfv\'enic within the simulation region.  
   For this component the energy flow is carried by the
 Poynting flux and the angular moment flow is carried by the
 magnetic field.
   
 \noindent{\bf Collimation and Driving of the Outflows:}
 Figure 4 shows the long-distance development of a conical wind from a slowly rotating star.   At large distances
the conical wind becomes collimated.

   To understand the collimation we analyzed
total force (per unit mass)
perpendicular to a poloidal magnetic field line (Lii et al. 2012).
For distances beyond the Alfv\'en surface of the flow this
force is approxiamtely
\begin{equation}
f_{\rm tot, \perp}  = - v_{\rm p}^2 \frac{\partial \Theta}{\partial s}
 - \frac{1}{8\pi \rho}\frac{\partial{\bf B}_p^2}{\partial n}
 -\frac{1}{8\pi\rho (r\sin\theta)^2}\frac{\partial(r\sin\theta B_\phi)^2}{\partial n} + \frac{v_\phi^2}{r}\frac{\cos\Theta}{\sin\theta}. 
\end{equation}
(Ustyugova et al. 1999).  Here, $\Theta$ is the angle between the poloidal magnetic field and the symmetry axis, $s$ is the arc length
along the poloidal field line, $n$ is a coordinate normal to
the poloidal field,  and the $p-$subscripts indicate
the poloidal component of a vector.
Once the jet begins to collimate, the curvature term $-v_{\rm p}^2\partial \Theta/\partial s$ also becomes negligible. 
      The magnetic force may act to either collimate or decollimate the jet, depending on the relative magnitudes of the toroidal $(r\sin\theta B_\phi)^2$ gradient (which collimates the outflow) and poloidal ${\mathbf B}_p^2$ gradient (which ``decollimates''). 
        In our simulations, the collimation of the matter implies that the magnetic hoop stress is larger than the poloidal field gradient. Thus the main perpendicular forces acting in the jet are the collimating effect of the toroidal magnetic field and the decollimating effect of the centrifugal force
and the gradient of ${\bf B}_p^2$.  The collimated effect of $B_\phi$ dominates.   Note that in MKS units  $2\pi r\sin\theta B_\phi/\mu_0$ is the poloidal current flowing through a surface of radius $r$ from colatitude zero to $\theta$.  For the jets from young stars this current is of the order of
$2 \times 10^{13}$ A.

\begin{figure*}[b]
\centering
\includegraphics[width=5.in]{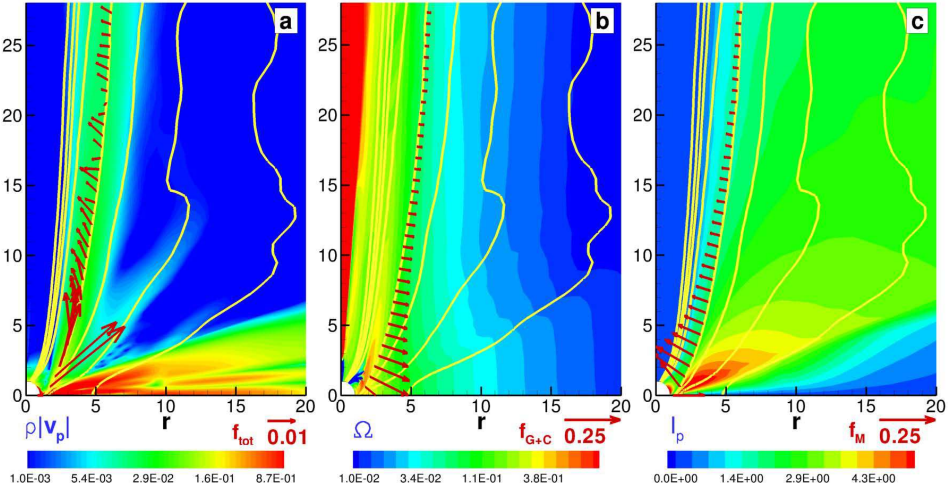}
\caption{Forces along a field line in the jet adapted from Lii 
et al. (2012).
{\bf Panel (a)} shows the poloidal matter flux density \rhov\ as a background overplotted with poloidal magnetic field lines. The vectors show the {\it total} force ${\mathbf f}_{\rm tot}$ along a representative field line originating from the disk at $r = 1.24$. 
{\bf Panel (b)} plots the angular velocity $\Omega$ as the background. The vectors show the sum of the {\it gravitational + centrifugal} forces ${\mathbf f}_{\rm G+C}$ along the representative field line.
{\bf Panel (c)} shows the poloidal current $I_p$ as the background. The vectors show the total {\it magnetic} force ${\bf f}_{\rm M}$ along the representative field line. }
\end{figure*}

        The driving force for the outflow
is simply the force parallel to the poloidal magnetic field of the flow $f_{{\rm tot},\parallel}$.
  This is obtained by taking the dot product of the Euler equation with the $\hat{\mathbf b}$ unit vector which is parallel to the poloidal magnetic field line ${\mathbf B}_p$.    The derivation by Ustyugova et al. (1999) gives
\begin{equation}
f_{\rm tot, \parallel}  
 =  - \frac{1}{\rho}\frac{\partial P}{\partial s} - \frac{\partial \Phi}{\partial s} + \frac{v_\phi^2}{r\sin\theta} \sin\Theta + \frac{1}{4 \pi \rho} \hat{\mathbf b} \cdot [(\nabla \times \bvec) \times \bvec]. 
\end{equation}
Here,  the terms on the right-hand side correspond to the
 pressure, gravitational, centrifugal and magnetic forces, respectively
 denoted ${\bf f}_{\rm P, G, C, M}$.
      The pressure gradient force, ${\bf f}_{\rm P}$, dominates within the disk. The matter in the disk is approximately in Keplerian rotation such that the sum of the gravitational and centrifugal forces roughly cancel (${\bf f}_{\rm G+C} \approx 0$). Near the slowly rotating star, however, the matter is strongly coupled to the stellar magnetic field and the disk orbits at sub-Keplerian speeds, giving ${\bf f}_{\rm G+C} \lesssim 0$.
    The  magnetic 
 driving force (the last term of Eq. 6) can be expanded as 
\begin{equation}
f_{\rm M, \parallel}  
 = -\frac{1}{8 \pi \rho (r\sin\theta)^2}{\partial \over \partial s}
 (r\sin\theta B_\phi)^2~,
\end{equation}
(Lovelace et al. 1991).

    Figure 5 shows the variation of the total force ${\bf f}_{\rm tot}$,
 the gravitational plus centrifugal force, and the magnetic force
 along a representative field line.
 This analysis establishes that the predominant driving force for the outflow is the magnetic force (Eq. 7) and not the centrifugal force.
   This in agreement with the analysis of Lovelace et al. (1991).

  {\bf Variability:}   For both rapidly and slowly
rotating stars     the magnetic field lines 
connecting the disc and the star have the tendency to
inflate and open (Lovelace,  Romanova \& Bisnovatyi-Kogan 1995).
Quasi-periodic reconstruction of the magnetosphere due to
inflation and reconnection has been discussed theoretically (Aly
\& Kuijpers 1990) and has been
observed in a number of axisymmetrtic simulations (Hirose et al.
1997; Goodson et al. 1997, 1999; Matt et al. 2002; Romanova et al.
2002).  Goodson \& Winglee (1999) discuss the physics of inflation cycles.
They have shown that each cycle of inflation 
consists of a period of matter accumulation near
the magnetosphere, diffusion of this matter through the magnetospheric field,
inflation of the corresponding field lines,
accretion of some matter onto the star, and outflow of some matter as winds,
 with subsequent expansion of the magnetosphere. There simulations show $5-6$
 cycles of inflation and reconnection.
 Our simulations often show $30-50$ cycles of inflation and reconnection.
    Figure 6 shows the time evolution of the accretion rates
for a of a slowly rotating star (Romanova et al. 2009).

\begin{figure*}[h]
\centering
\includegraphics[width=3.in]{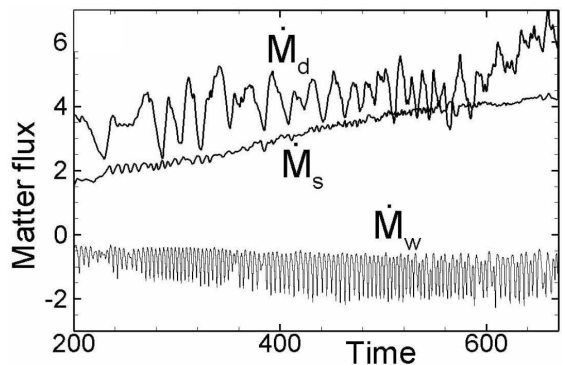}
\caption{Matter flux onto the star $\dot M_s$, into the
conical wind $\dot M_w$, and through the disk $\dot M_d$. }
\end{figure*}

\begin{figure*}[h]
\centering
\includegraphics[width=4.in]{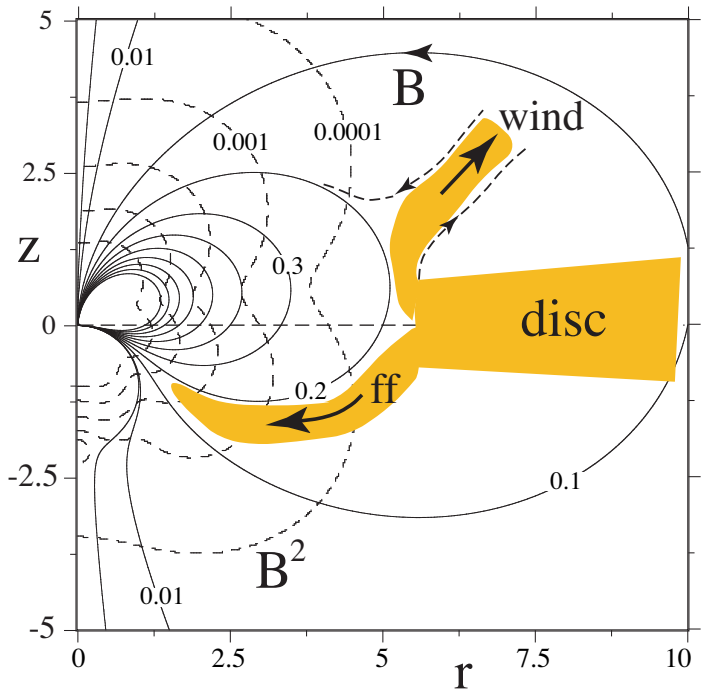}
\caption{The initial poloidal magnetic field lines 
and constant magnetic pressure lines for
the case of an aligned dipole and quadrupole field adapted from
Lovelace et al. (2010).
The funnel flow (ff) and the wind in this figure are suggested.
The dashed lines are constant values of ${\bf B}^2$.}
\end{figure*}

Kurosawa and Romanova (2012) have calculated spectra from modeled conical winds using the radiative transfer code TORUS
and have shown that conical winds may explain different features in the hydrogen spectral lines,  in the He I line and
also a relatively narrow, low-velocity blue-shifted absorption 
components in the He I $\lambda 10830$ which is often seen in observations (Kurosawa et al. 2011).

\section{One-Sided and Lop-Sided Jets}

There is clear evidence,  mainly from
Hubble Space Telescope (HST) observations,  of the asymmetry
between the approaching and receding jets
from a number of young stars.
The objects include the jets in HH 30 (Bacciotti et al. 1999),
RW Aur (Woitas et al. 2002), TH 28 (Coffey et al. 2004),
and LkH$\alpha$ 233  (Perrin \& Graham 2007).
  Specifically, the radial speed of the approaching
jet may differ by a factor of two from that of
the receding jet.
    For example, for RW Aur the radial redshifted
speed is $\sim 100$ km/s whereas the blueshifted
radial speed is $\sim175$ km/s.
   The mass and momentum fluxes are also
significantly different
for the approaching and receding jets in a number of cases.
  Of course, it is possible that the observed asymmetry
of the jets could be due to say differences in the gas densities
on the two sides of the source.   Here, we discuss
the case of intrinsic asymmetry where the asymmetry of outflows is
connected with asymmetry of the star's magnetic field.
      Substantial  observational evidence points to the fact that
young stars often have {\it complex} magnetic  fields consisting
of    dipole,  quadrupole, and higher order  poles 
misaligned with respect to each other and the rotation
axis (Jardine et al. 2002; Donati et al.  2008).
Analysis of the plasma flow around stars with realistic fields
have shown that a fraction
of the star's magnetic field lines are open
and may carry outflows (e.g., Gregory et al. 2006).

        It is evident that the  complex magnetic field 
of a star will destroy the commonly assumed symmetry of the magnetic field and the plasma about the equatorial plane.
   Figure 7 shows an illustrative complex magnetic field consisting of
the combination of a dipole and a quadrupole field both of which are axisymmetric.
The figure includes the suggested locations of the funnel 
flow to the star (Romanova et  al. 2002) and the conical wind outflows. 
   The MHD simulations fully support the qualitative picture
suggested in Figure 7 (Lovelace et al. 2010).  The time-scale during
which the jet comes from the upper hemisphere is set
by the evolution time-scale for the stellar magnetic field.
This is determined by the dynamo processes responsible for
the generation of the field. 

      Remarkably, once the assumption of symmetry about the
equatorial plane is dropped, we find that the conical winds
alternately come from one hemisphere and then the other 
{\it even} when the stellar magnetic field is a centered
axisymmetric dipole  (Lovelace et al. 2010).  
An illustrative case of this spontaneous symmetry breaking is shown
in Figure 8.  The time-scale for the `flipping' is the accretion time-scale
of the inner part of the disc which is expected to be much less
than the evolution time of the star's magnetic field.

\begin{figure*}[t]
\centering
\includegraphics[width=5.in]{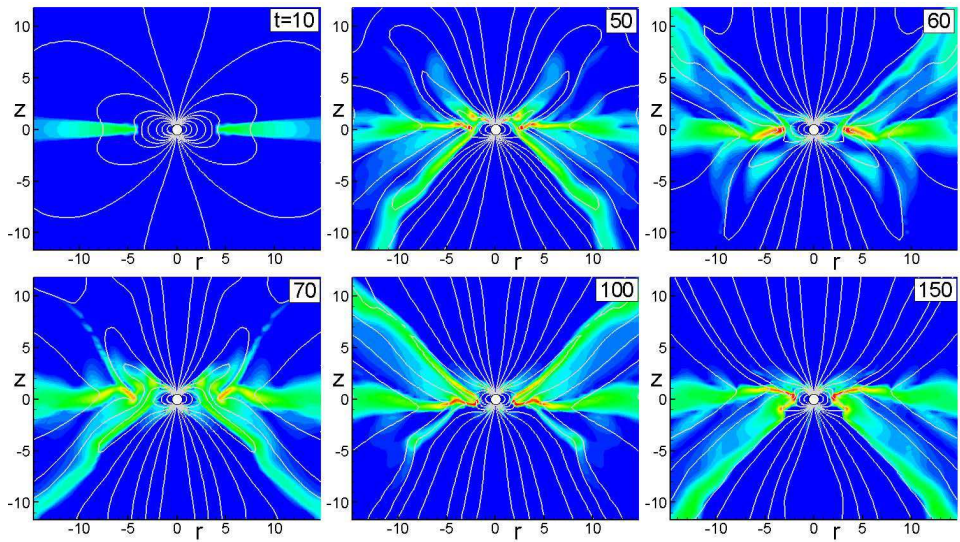}
\caption{"Flip-flop"of outflows in the case 
where the stellar magnetic field is a
centered axisymmetric dipole adapted from Lovelace
et al. (2010).
The color background shows the matter flux
distribution, and the lines are the poloidal magnetic field lines.}
\end{figure*}

\section{Conclusions}

      Detailed magnetohydrodynamic simulations have
established that long-lasting outflows of  cold disc matter into a hot low-density corona from the disc-magnetosphere boundary in cases of slowly and rapidly rotating stars.  The main results are the following:

  For slowly rotating stars 
a new type of outflow ---  a conical wind --- has been discovered.
  Matter flows out forming a {\it conical wind} which has the shape of a thin conical shell with a half-opening angle $\theta \sim 30^\circ$.
The outflows appear in cases where the magnetic flux of the star is bunched up by the inward accretion flow of the disc.
We find that this occurs when the turbulent magnetic Prandtl number
(the ratio of viscosity to diffusivity)  ${\cal P}_m > 1$, and when the viscosity is sufficiently high, $\alpha_\nu\gtrsim 0.03$.

Winds from the disc-magnetosphere boundary
have been proposed earlier by Shu and collaborators and referred to as
X-winds (Shu et al. 1994).
In this model,  the wind originates from a small region near the corotation radius $r_{\rm cr}$,
while the disc truncation radius $r_t$ (or, the magnetospheric radius $r_{\rm m}$) is only slightly smaller than
$r_{\rm cr}$ ($r_{\rm m}\approx 0.7 r_{\rm cr}$, Shu et al. 1994).
   It is suggested that excess angular
momentum flows  from the star to the disc and from there into the X-winds.
  The model aims to explain the slow rotation of the star and the formation of jets.
In the simulations discussed here we have  obtained outflows from both
slowly and rapidly rotating stars.
Both have conical wind components which are reminiscent of  X-winds.
     In some respects  the conical winds are similar to
X-winds:  They both require {\it bunching} of the poloidal field lines and show outflows from the inner disc; and they both have high rotation and show gradual poloidal acceleration (e.g., Najita \& Shu 1994).

       The main differences are the following:
{(1)} The conical/propeller outflows have {\it two components}: a slow high-density  conical wind (which can be considered as an analogue of the X-wind), and a fast low-density jet.
No jet component is discussed in the X-wind model.
{(2)} Conical winds form around stars with {\it any rotation rate} including very slowly rotating stars. They do not require fine tuning of the corotation and truncation radii.
For example, bunching  of field lines is often expected during periods of enhanced or unstable accretion when the disc comes closer to the surface of
the star and  $r_{\rm m}<<r_{\rm cr}$. Under this condition conical winds will form. In contrast, X-winds require $r_{\rm m}\approx r_{\rm cr}$.~ { (3)}
The base of the conical wind component in both slowly and rapidly rotating stars is associated with the region where the field lines are bunched up, and not with the corotation radius.
{(4)} X-winds are driven by the {\it centrifugal force}, and as a result matter flows over a wide range of directions below the ``dead zone"   (Shu et al. 1994; Ostriker \& Shu 1995).
In conical winds the matter is driven by the {\it magnetic force} (Lovelace et al. 1991) which acts such that
the matter flows into a {\it thin shell} with a cone half-angle $\theta\sim 30^\circ$. The same force tends to  collimate the flow.

   For rapidly rotating stars in the propeller regime where
$r_{\rm m} > r_{\rm cr}$ and where the condition for bunching, ${\cal P}_m>1$, is satisfied
we find two distinct outflow components (1) a relatively low-velocity conical wind and (2) a high-velocity axial jet.
 A significant part of the disc matter and  angular momentum flows into the conical winds.
   At the same time a significant part of the rotational energy of the star flows
into the magnetically-dominated axial jet.
   This regime is particularly relevant to protostars, where the star rotates rapidly and has a high accretion rate.
   The star spins down rapidly due to the angular momentum flow into the axial jet along the field lines connecting the star and the corona. For typical parameters a protostar spins down in $3\times10^5$ years. The axial jet is powered by the spin-down of the star rather than by disc accretion.
    The matter fluxes into both components (wind and jet) strongly oscillate due to events of inflation and reconnection. Most powerful outbursts occur every $1-2$ months. The interval between outbursts is expected to be longer for smaller diffusivities in the disc.
Outbursts are accompanied by higher outflow velocities and stronger self-collimation of both components. Such outbursts may
explain the ejection of knots in some CTTSs every few months.

      When the artificial requirement of symmetry about the 
equatorial plane is dropped,  MHD simulations reveal
that  the conical winds may 
alternately come from one side of the disc and then the other
even for the case where the stellar magnetic field is a
centered axisymmetric dipole (Lovelace et al. 2010).

   In recent work we have studied the disc accretion
to rotating magnetized stars in the propeller regime 
using a new code with very high resolution in the region
of the disc (Lii et al. 2013).  
In this code  {\it no} turbulent viscosity or diffusivity is incorporated,
but instead strong turbulence occurs due to the magneto-rotational
instability.  This turbulence drives the accretion and it leads
to  episodic outflows.

The authors thank G.~V. Ustyugova and A.~V. Koldoba for the development of the codes used in the
reported simulations.   This research was supported in part by NSF grants AST-1008636 and AST-1211318 and by a NASA ATP grant NNX10AF63G; we thank NASA for use of the NASA High Performance Computing Facilities.

\section*{References}
\begin{harvard}

\item[] Alpar, M.A., \& Shaham, J. 1985, Nature, 316, 239

\item[] Aly, J.J., \& Kuijpers, J. 1990, A\&A, 227, 473

\item[] Bacciotti, F., Eisloffel, J., \& Ray, T.P. 1999,
A\&A, 350, 917

\item[] Bessolaz, N., Zanni, C., Ferreira, J., Keppens, R., Bouvier,
J. 2008, A\&A, 478, 155

\item[] Blandford, R.D.., \& Payne, D.G. 1982, MNRAS, 199, 883

\item[] Cabrit, S., Edwards, S., Strom, S.E., \& Strom, K.M.
1990, ApJ, 354, 687

\item[] Cai, M.J., Shang, H., Lin, H.-H., \& Shu, F.H. 2008, ApJ,
672, 489

\item[] Casse, F., \& Keppens, R. 2004, ApJ, 601, 90

\item[] Coffey, D., Bacciotti, F., Woitas, J., Ray, T.P., \&
Eisl\"offel, J. 2004, ApJ, 604, 758

\item[] Donati, J.-F., Jardine, M. M., Gregory, S. G., Petit, P., Paletou, F.,
Bouvier, J., Dougados, C., M\`{e}nard, F., Cameron, A. C., Harries, T. J.,
Hussain, G. A. J., Unruh, Y., Morin, J., Marsden, S. C., Manset, N., Auri\`{e}re, M., Catala, C., Alecian, E. 2008, MNRAS, 386, 1234

\item[] Edwards, S., Fischer, W., Hillenbrand, L., Kwan, J. 2006,
ApJ, 646, 319

\item[] Edwards, S., Fischer, W., Kwan, J., Hillenbrandt, L.,
Durpee, A.K. 2003, ApJ, 599, L41

\item[] Edwards, S. 2009, Proceedings of the 15th Cambridge Workshop on Cool Stars, Stellar Systems and the Sun. AIP Conference Proceedings, Volume 1094, pp. 29-38

\item[] Ferreira, J, Dougados, C., \& Cabrit, S. 2006, A\&A, 453,
785

\item[] Goodson, A.P., \& Winglee, R. M., 1999, ApJ, 524, 159

\item[] Goodson, A.P., Winglee, R. M., \& B\"ohm, K.-H. 1997, ApJ,
489, 199

\item[] Goodson, A.P., B\"ohm, K.-H., Winglee, R. M. 1999, ApJ, 524,
142

\item[] Gregory S. G., Jardine M., Simpson I., \& Donati J.-F., 2006, MNRAS, 371, 999

\item[] Hartigan, P, Edwards, S., \& Gandhour, L. 1995, ApJ, 452, 736

\item[] Hayashi, M. R., Shibata, K., \& Matsumoto, R. 1996, 468, L37

\item[] Heinz, S., Schulz, N. S., Brandt, W. N., \& Galloway, D. K.
2007, ApJ, 663, L93

\item[] Hirose, S., Uchida, Yu.,  Shibata, K., \& Matsumoto, R. 1997,
PASJ, 49, 193

\item[] Hsu, S.C. \& Bellan, P.M. 2002, MNRAS, 334, 257

\item[] Illarionov, A.F., \& Sunyaev, R.A. 1975, A\&A, 39, 185

\item[] Jardine, M., Collier Cameron, A., \& Donati, J.-F. 2002,
MNRAS, 333, 339

\item[] K\"onigl, A., \& Pudritz, R. E. 2000, in {\it Protostars and Planets
IV}, Mannings, V., Boss, A.P., Russell, S. S. (eds.), University of
Arizona Press, Tucson, p. 759

\item[] K\"uker, M., Henning, T., \& R\"udiger, G. 2003, ApJ, 589, 397

\item[] Kurosawa, R., Romanova, M. M., \& Harries, T.  2011, MNRAS, 416, 2623

\item[] Kurosawa, R., \& Romanova, M. M. 2012, MNRAS, 426, 2901

\item[] Kwan, J., Edwards, S., \& Fischer, W. 2007, ApJ, 657, 897

\item[] Lebedev, S. V., Ciardi, A., Ampleford, D. J., Bland, S. N., Bott, S. C., Chittenden, J. P., Hall, G. N.,  Rapley, J.,  Jennings, C. A.,  Frank, A.,  Blackman, E. G., \& Lery, T. 2005, MNRAS, 361, 97

\item[] Lii, P.S., Romanova, M., \& Lovelace, R. 2012, 
MNRAS, 420, 2020

\item[] Lii, P.S., Romanova, M.M., Ustyugova, G.V., Koldoba, A.V.,
\& Lovelace, R.V.E. 2013, MNRAS, in press (arXiv: 1304.2703v1)

\item[] Lovelace, R.V.E., Berk, H.L., \& Contopoulos, J. 1991, ApJ, 379,
696

\item[] Lovelace, R.V.E., Romanova, M.M., \& Bisnovatyi-Kogan, G.S. 1995, MNRAS, 275, 244

\item[] Lovelace, R.V.E., Romanova, M.M., \& Bisnovatyi-Kogan, G.S. 1999, ApJ, 514, 368

\item[] Lovelace, R. V. E.,  Romanova, M. M.,  Ustyugova, G. V., \&  Koldoba, A. V. 2010, MNRAS, 408, 2083

\item[] Matt, S., Goodson, A.P., Winglee, R.M., \& B\"ohm, K.-H. 2002, ApJ, 574, 232

\item[] Miller, K.A., \& Stone, J.M. 1997, ApJ, 489, 890

\item[] Najita, J.R., \& Shu, F.H. 1994, ApJ, 429, 808

\item[] Ostriker, E.C., \& Shu, F.H. 1995, ApJ, 447, 813

\item[] Perrin, M.D., \& Graham, J.R. 2007, ApJ, 670, 499

\item[] Romanova, M.M., Ustyugova, G.V., Koldoba, A.V., \&
Lovelace, R.V.E. 2002, ApJ, 578, 420

\item[] Romanova, M.M., Ustyugova, G.V., Koldoba, A.V., \&
Lovelace, R.V.E. 2009, MNRAS, 399, 1802

\item[] Shakura, N.I., \& Sunyaev, R.A. 1973, A\&A, 24, 337

\item[] Shu, F., Najita, J., Ostriker, E., Wilkin, F., Ruden, S.,
Lizano, S. 1994, ApJ, 429, 781

\item[] Shu, F.H, Galli, D., Lizano, S., Glassgold, A.E., \&
Diamond, P.H. 2007, ApJ, 665, 535

\item[] Sokoloski, J.L., \& Kenyon, S.J. 2003, ApJ, 584, 1021

\item[] Ustyugova, G.V., Koldoba, A.V., Romanova, M.M., Chechetkin,
V.M., \& Lovelace, R.V.E. 1999, ApJ, 516, 221

\item[] Ustyugova, G.V., Koldoba, A.V., Romanova, M.M., \& Lovelace, R.V.E. 2006, ApJ, 646, 304 

\item[] Woitas, J., Ray, T.P., Bacciotti, F., Davis, C.J., \& Eisl\"offel, J.
2002, ApJ, 580, 336

\end{harvard}

\end{document}